# The Anion Effect on $Li^+$ Ion Coordination Structure in Ethylene Carbonate Solutions

Bo Jiang[1,3#], Veerapandian Ponnuchamy[4,5#*,] , Yuneng Shen[1#], Xueming Yang[1], Kaijun Yuan[1*], Valentina Vetere[5], Stefano Mossa[4], Ioannis Skarmoutsos[4*], Yufan Zhang[3], Junrong Zheng[1,2,3]*

[1]*State key Laboratory of Molecular Reaction Dynamics, Dalian Institute of Chemical Physics, Chinese Academy of Sciences, 457 Zhongshan Road, Dalian 116023, China*

[2]*College of Chemistry and Molecular Engineering, Beijing National Laboratory for Molecular Sciences, Peking University, Beijing 100871, China*

[3]*Department of Chemistry, Rice University, Houston, TX 77005*

[4]*Univ. Grenoble Alpes, INAC-SYMMES, F-38000 Grenoble, France; CNRS, INAC-SYMMES, F-38000 Grenoble, France; CEA, INAC-SYMMES, F-3800 Grenoble, France*

[5]*Univ. Grenoble Alpes, LITEN-DEHT, F-38000 Grenoble, France and CEA, LITEN-DEHT, F-38000 Grenoble, France*

** To whom correspondence should be addressed. E-mails: kjyuan@dicp.ac.cn; zhengjunrong@gmail.com; veera.pandi33@gmail.com ; iskarmoutsos@hotmail.com*

*# These authors have equally contributed to this work*

# Abstract

Rechargeable lithium ion batteries are an attractive alternative power source for a wide variety of applications. To optimize their performances, a complete description of the solvation properties of the ion in the electrolyte is crucial. A comprehensive understanding at the nano-scale of the solvation structure of Lithium ions in non-aqueous carbonate electrolytes is, however, still unclear. We have measured by femtosecond vibrational spectroscopy the orientational correlation time of the CO stretching mode of $Li^+$-bound and $Li^+$-unbound ethylene carbonate molecules, in $LiBF_4$, $LiPF_6$ and $LiClO_4$ ethylene carbonate solutions with different concentrations. Surprisingly, we have found that the coordination number of ethylene carbonate in the first solvation shell of $Li^+$ is only two, in all solutions with concentrations higher than 0.5 M. Density functional theory calculations indicate that the presence of anions in the first coordination shell modifies the generally accepted tetrahedral structure of the complex, allowing two EC molecules only to coordinate to $Li^+$ directly. Our results demonstrate for the first time, to the best of our knowledge, the anion influence on the overall structure of the first solvation shell of the $Li^+$ ion. The formation of such a cation/solvent/anion complex provides a rational explanation for the ionic conductivity drop of lithium/carbonate electrolyte solutions at high concentrations.

Lithium ion batteries (LIBs) are extensively used in small electronic devices such as cell phones and notebook computers. They are also growing in popularity for automotive applications, in order to decrease the greenhouse gas emissions and air pollution. The typical lithium-ion battery is composed by two electrodes which intercalate lithium materials, and a layer of a non-aqueous electrolyte solution, which directly determines the conductivity of the device[1-3]. The dynamics of $Li^+$ through the electrolyte controls the rate of the energy transfer. It is generally accepted that transport of $Li^+$ ions can be described as a complex mechanism, involving first solvation of the ions by the solvent molecules, followed by the migration of the solvated ions[4]. The ionic conductivity in a solution is, as a consequence, the overall result of these two steps.

In the last two decades, an impressive amount of theoretical and experimental work has focused on the interactions of $Li^+$ with carbonate-based electrolytes that are applied in batteries. In particular, spectroscopy techniques, such as Nuclear Magnetic Resonance (NMR), Raman Spectroscopy, and Electrospray Ionisation Mass Spectrometry (ESI-MS), have been employed to determine the coordination numbers corresponding to the first solvation shell of the ion. The details of the solvation structure and of the dynamics of lithium ions in these solvents are, however, still a subject of lively debate. NMR measurements indicate that up to six ethylene carbonate molecules can coexist in the $Li^+$ solvation shell[5-7]. Raman spectroscopy measurements[8] concluded similarly as the NMR results. Intriguingly, MS measurements point in the opposite direction, suggesting much lower coordination numbers[9]. We add that none of these techniques has been able to provide detailed information about the modifications induced by the presence of the anions, or on the solvation structure under conditions sufficiently similar to those of the actual environments found in batteries. In the MS experiments, the solvated $Li^+$ ion was studied in the gaseous state, and probably already experienced partial dissociation before being observed. In the NMR experiments, the $^{13}C$ or $^{17}O$ nuclei of the solvent molecules provide only an indirect probe of $Li^+$ solvation, and it is therefore difficult to derive quantitative information from the associated chemical shifts. In linear vibrational spectroscopy measurements, the derivation of coordination number suffers from the unknown ratio of transition dipole moments of $Li^+$-bound and unbound solvent molecules and the unknown salt dissociation constant[10].

Numerical studies were not conclusive either. Both ab-initio calculations and Molecular Dynamics (MD) simulations suggest that the small ionic radius of lithium cannot directly bind more than five organic solvent molecules, and the complex with a (total) coordination number of four is the most stable[11-16]. Unfortunately, both techniques also appear problematic in the present case. On one side, the relevance of DFT-stable finite-size clusters for the bulk electrolyte is not obvious. On the other, one must emphasize that the validity of the force fields used in MD simulations can only be scrutinized by a direct comparison with experimental data. The development of experimental methods proving a *direct* determination of the coordination number therefore becomes indispensable, for obtaining from simulation an unambiguous picture of the solvation properties of lithium ion solutions.

Here, we propose an integrated experimental spectroscopy / ab-initio calculations investigation (with some reference to our previous MD work[17]), in order to minimize the limitations of the two approaches when used separately. We have employed a combination of nonlinear vibrational spectroscopy and DFT calculations to provide deeper insight on the solvation structure of lithium in solutions of high concentrations similar to those in $Li^+$ batteries. Our studies suggest a surprising $Li^+$ solvation structure sensibly different from the generally accepted tetrahedral coordination of the carbonyl oxygen atoms around $Li^+$.

We have focused on the solvation structure of different $Li^+$ ion solutions, and measured the orientational correlation time of $Li^+$-bound and $Li^+$-unbound (free) solvent molecules for different salt solutions, by using infrared femtosecond spectroscopy. All details of the methodology are described in our previous publication[10] and also given in the Supporting Information (SI). Here we only recall that with this method we can observe the dynamical behavior of the different solvent chemical species on time scales shorter than the typical time associated to the exchange mechanism(s) between the solvation shells and the bulk liquid. According to the Stokes-Einstein equation, $D_r = \frac{kT}{8\pi \cdot \eta \cdot r^3}$, the rotational diffusion time constant is proportional to the volume of the molecule[18]. Here, $D_r$ is the rotational diffusion time constant, η the viscosity, r is the particle radius, T the temperature, and $k$ is the Boltzmann constant.) As a consequence, the heavier the considered molecule, the slower the associated orientational decay time is. Here, we want to emphasize that the equation is rigorously valid only for ideal conditions which a real system can hardly achieve. In order to evaluate the applicability of the equation in determining molecular volumes in liquids, we measured three different systems: (1) a series of flexible molecules with different alkyl lengths; (2) rigid molecules with different numbers of benzene rings; and (3) $Li^+$ salts in $CH_3CN$ of which the crystalline structures of the $Li^+$-$CH_3CN$ complexes have been identified by XRD. Experimental results show that the method works well for all the three different systems[10]. Details of the experimental methods are provided in the Supporting information.

We show in Figure 1.A the absorption spectra of ethylene carbonate (EC) and $LiBF_4$/EC solutions at 1/5, 1/10 and 1/20 mol% concentration, in the frequency region of the O-C-O stretching mode. The main peak for the pure ethylene carbonate (black line) centered at 1165 $cm^{-1}$ is assigned to the O-C-O mode mentioned above. The secondary peak of small intensity at ~1222 $cm^{-1}$ is assigned to a combination band. With the addition of $LiBF_4$, the stretching and combination band features keep the same position, increasing in intensity. Interestingly, a third attribute develops at ~1203 $cm^{-1}$, and its intensity grows by increasing the concentration of $LiBF_4$. We associate this observation to the binding of either $Li^+$ or $BF_4^-$ with EC, causing the frequency blue shifts.

To unambiguously identify the ion responsible of this latter phenomenon, FTIR spectra of $LiClO_4$, $LiPF_6$, $NaClO_4$ and $KClO_4$, dissolved in EC at the indicated fixed concentration, are shown in Figure 1.B. From these data it is clear that a change in the nature of the anions (from $BF_4^-$ to $PF_6^-$ or $ClO_4^-$) does not influence the frequency position of the peak. In contrast, modifications of the cation (from $Li^+$ to $Na^+$ or $K^+$) even completely destroy the peak at 1203 $cm^{-1}$.

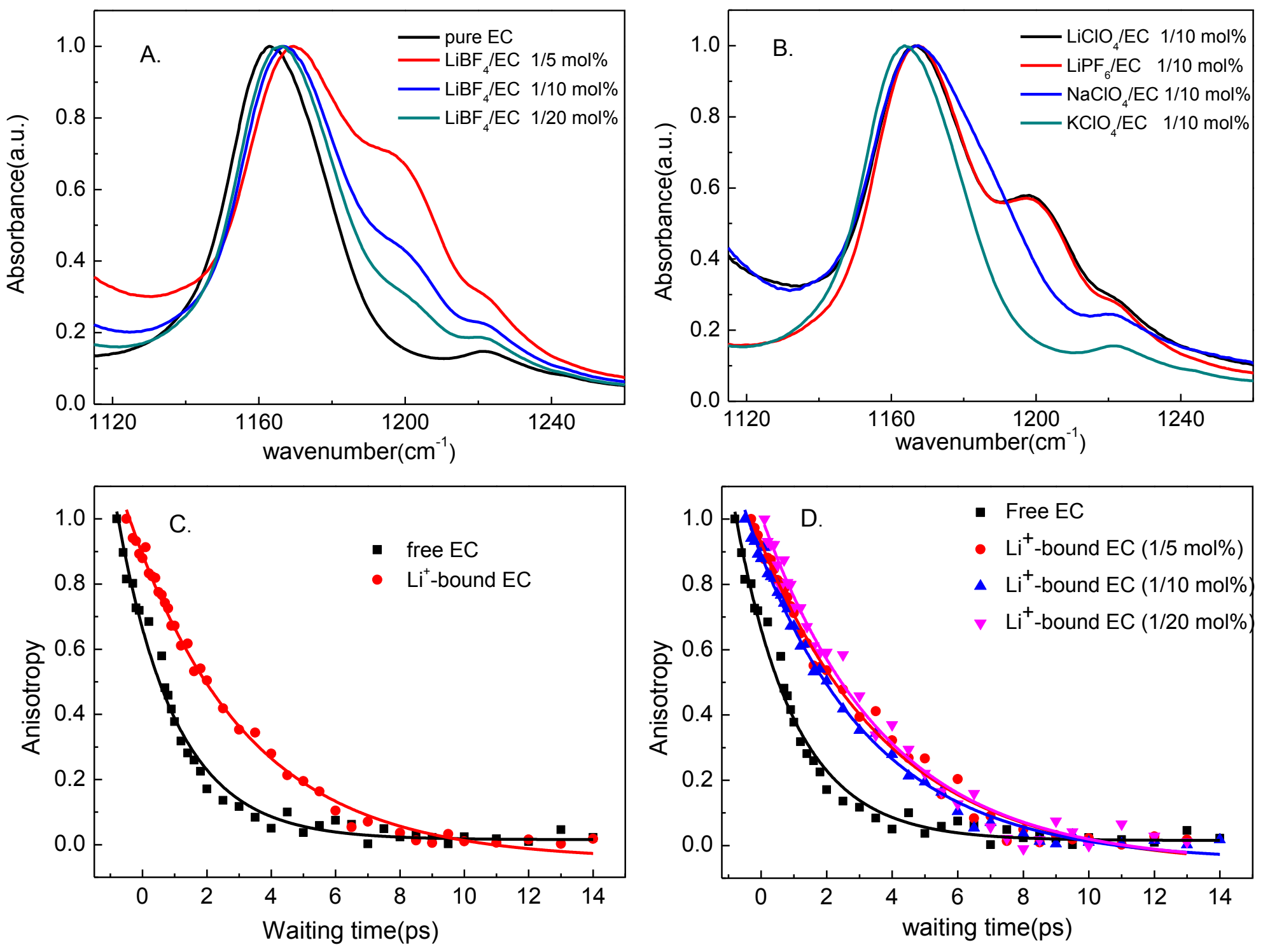

***Figure 1:*** A. FTIR spectra of ethylene carbonate (EC), 1/5 mol%, 1/10 mol%, 1/20 mol% $LiBF_4$/EC. B. FTIR spectra of EC, 1/10 mol% $LiClO_4$/EC, 1/10 mol% $LiPF_4$/EC, 1/10 mol% $NaClO_4$/EC, 1/10 mol% $KClO_4$/EC. C. Time dependent anisotropies of CO stretch of free EC (1165 cm$^{-1}$) and $Li^+$-bound EC (1203 cm$^{-1}$) in a 1/10 mol% $LiBF_4$/EC solution. D. Time dependent anisotropies of CO stretch of $Li^+$-bound EC (1203 cm$^{-1}$) with the concentration of 1/5, 1/10, 1/20 mol% $LiBF_4$/EC solution.

This conclusion has an important implication: the orientational dynamics of (free) bulk-like solvent molecules, and that of solvent molecules bound to the $Li^+$ ion can be studied separately. In Figure 1.C we show the measured time decay of the anisotropy parameter R, for the 1/10 mol% $LiBF_4$/EC solution. The decay time of parameter $R = \frac{(P_\parallel - P_\perp)}{(P_\parallel + 2P_\perp)}$ represents the orientational correlation time of the solvent molecules (where $P_\parallel$ and $P_\perp$ are the parallel and perpendicular signals, respectively). In these measurements, the O-C-O stretch band of the free EC (1165 cm$^{-1}$) and of the $Li^+$- bound EC (1203 cm$^{-1}$) was excited and detected at the 0→1 transition. Based on our previous work[10], the anisotropy decay is mainly controlled by the molecular rotation, since both resonant and non-resonant energy transfers are much slower than the molecular rotation in EC solutions (see the SI). Interestingly, for the $Li^+$-bound EC, the anisotropy decay is 3.7±0.2 ps, which is twice that of unbound EC (1.8 ± 0.1 ps). We will discuss the implications of this finding in what follows.

Next, in Figure 1.D we show the anisotropy decay curves of the stretch mode of $Li^+$-bound EC, in 1/5 mol%, 1/10 mol% and 1/20 mol% $LiBF_4$/EC solutions. The time constants extracted from the data are now 3.7±0.2 ps, 3.6±0.2 ps and 3.7±0.2 ps, respectively in the three cases. The decay of R therefore turns out to be independent of the salt concentration, implying that the probability for multiple anions to strongly bind to the cation or EC is very small. The results seem to be consistent with the conclusion from recent work[19] that "anions exist relatively free with little solvation". Indeed, further experimental evidence on additional solutions, including $LiPF_6$/EC and $LiClO_4$/EC, support the conclusion that the measured anisotropy decays are determined by the rotational time scale of complexes without any binding to multiple anions (see Table 1 and Figures S2-6). We have performed a similar analysis for all investigated lithium salts, obtaining the rotational time ratios for the $Li^+$/EC complex over the free EC molecule (in the same solutions to remove the overall viscosity effect) listed in Table 1, at the indicated concentrations. In all solutions, the values of the rotation ratio are comprised in the range 2.0 to 2.3. Here an issue is worth noting. In our previous work[10], we developed a method to quantitatively measure the dissociation constants of $Li^+$ salts in organic solvents. Our results are consistent with the well-known knowledge that $LiPF_6$ has a larger dissociation constant than $LiBF_4$ in $CH_3CN$ or EC and with increasing salt concentration more cation-anion ion pairs form. However, what our experiments measure here is the $Li^+$-bound EC. With the increase of salt concentration, the salt dissociation/association equilibrium may shift to more cation/anion ion pair from EC solvated $Li^+$, but such an equilibrium shift does not necessarily change the solvation structure of $Li^+$. We believe that this is the intrinsic reason for that our results are insensitive to anion and salt concentrations.

***Table 1:*** The time constants of anisotropy decay of free EC and $Li^+$-bound EC in the Lithium salts/EC solutions with different concentrations. All measurements were performed at 40 $^0$C.

| **Rotation time constant** | **Free EC (ps)** | **Li+-bound EC (ps)** | **Rotation ratio** |
|---|---|---|---|
| ***LiBF₄/EC*** (1/5 mol%) | 1.8±0.1 | 3.7±0.2 | 2.1 |
| (1/10 mol%) | 1.8±0.1 | 3.6±0.2 | 2.0 |
| (1/20 mol%) | 1.8±0.1 | 3.7±0.2 | 2.1 |
| ***LiPF₆/EC*** (1/10 mol%) | 1.9±0.1 | 4.3±0.2 | 2.3 |
| (1/20 mol%) | 2.0±0.1 | 4.4±0.2 | 2.2 |
| ***LiClO₄/EC*** (1/5 mol%) | 1.9±0.1 | 4.1±0.2 | 2.1 |
| (1/10 mol%) | 1.9±0.1 | 4.3±0.2 | 2.3 |
| (1/20 mol%) | 1.8±0.1 | 4.0±0.2 | 2.2 |

Interestingly, we are now in the position to directly associate these (dynamical) results to the detailed (structural) composition of the detected rotating units. Indeed, based on the Stokes-Einstein equation introduced above, this finding indicates that the volume of the $Li^+$/EC complex is about two times bigger than that pertaining to one free EC molecule. Excluding the negligible volume occupied by $Li^+$, we can therefore immediately conclude that two EC molecules bind to $Li^+$ in the observed complexes, a value sensibly lower than the total coordination number of four reported by previous experimental and theoretical studies on solutions[14]. We also note that, using the same method above, a coordination number of four was determined for $Li^+$ in $CH_3CN$ solutions, which is in agreement with X-Ray diffraction measurements[10].

To further explore the above surprising experimental results, we have performed DFT numerical calculations. We have used the Amsterdam Density Functional (ADF) software[20], and considered the PBE GGA functional and a TZP (core double zeta, valence triple zeta, polarized) slater type orbital (STO) basis set for the geometry optimization. Additional calculations for a few representative clusters have confirmed that the basis set superposition error (BSSE) corrections are negligibly small, a conclusion corroborated by our recent DFT studies[17]. Frequency analysis has been performed for each structure, ensuring the absence of negative eigenvalues to confirm them as true minima of the potential energy surface. Zero-point energy (ZPE) corrections have been also taken into account. Thermodynamic potentials associated to the investigated clusters have been determined in the gas-phase approximation and at temperature T=298.15 K. The calculation temperature is slightly different from the experimental temperature (40 $^0$C). However, since the free energy difference caused by the small temperature difference is much smaller than 2 kcal/mol and the binding energy of the $Li^+$ complex with two EC molecules is higher than other complexes for at least 2 kcal/mol, the calculation/experiment temperature difference doesn't affect our conclusion. Additional details can be found in the SI.

We have first performed the gas-phase DFT energy optimizations for the $Li^+(EC)_{1\text{-}3}$(Anion-) clusters. Following the methods employed in our previous study[17], enthalpy, entropy and Gibbs free-energy were also calculated, in order to estimate the preferential structure of the first coordination shell around the lithium cation, in the presence of the counter-anion. More in details, we have calculated the free energy changes for clusters of the form $Li^+(S)_n$(Anion-) + m S (here S is the solvent species, n + m = 3, and Anion- is the counter-anion, $ClO_4^-$, $BF_4^-$ or $PF_6^-$). Subsequently, we have determined the free energy differences associated to the transitions:

$$Li^+(S)_k\,Anion^- + l\cdot S \rightarrow Li^+(S)_n\,Anion^- + m\cdot S \;,\; k+l=n+m,\; k,l,n,m=0-3$$

In cases where the above transitions exhibit negative free energy changes, this approach provides a clear indication that the formation of the $Li^+(S)_n$(Anion-) cluster is favourable. The obtained free energy changes, shown in Table 2, reveal that clusters including two solvent molecules and one counter-ion around the $Li^+$ ion are the most thermodynamically stable. The optimized most stable structures are presented in Figure 2, for each case. In order to also test if the trends observed for gas-phase calculations change upon the addition of a dielectric continuum, we used the COSMO (COnductor-like Screening Model) model[21] to approximate solvent effects. Our tests, using the dielectric constant of EC for the continuum, have shown that the clusters containing one anion and two EC molecules around lithium are the most thermodynamically stable even when a dielectric continuum is added (see Table S1 in the SI Section).

Two important observations are now in order. First, in our earlier DFT study we have demonstrated that the most stable clusters in diluted conditions contain four solvent molecules around the $Li^+$ ion[17], organized with a strong tetrahedral symmetry. Here, in contrast, we find that the addition of the counter-anion both modifies the overall coordination number, which now is three instead of four, and destroys the local (tetrahedral) symmetry.

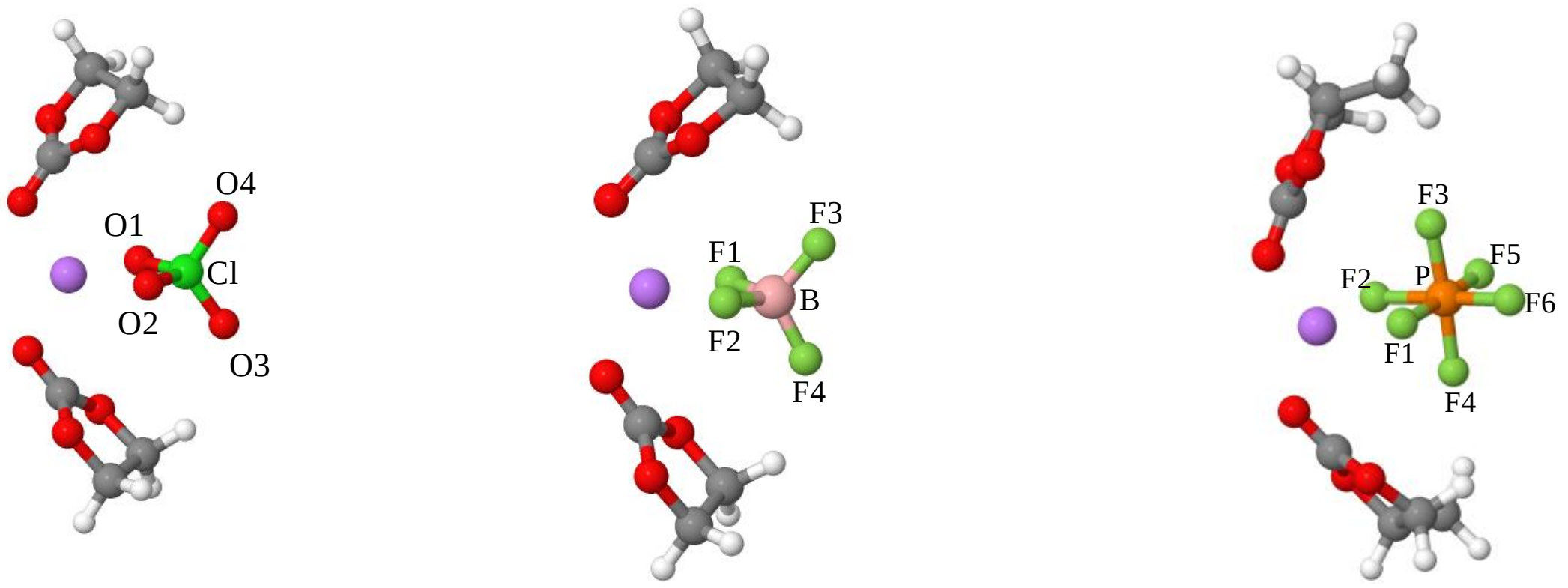

(a) (b) (c)

***Figure 2:*** The optimized geometries of the most stable clusters: (a) $Li^+(EC)_2ClO_4^-$, (b) $Li^+(EC)_2BF_4^-$ and (c) $Li^+(EC)_2PF_6^-$.

***Table 2:*** Gibbs free energy of Clusters $Li^+(EC)_{0\text{-}3}$Anion- (where Anion- = $ClO_4^-$, $BF_4^-$, $PF_6^-$) (kcal/mol)

| **Fragments** | **G** | | |
|---|---|---|---|
| | Anion- = $ClO_4^-$ | Anion- = $BF_4^-$ | Anion- = $PF_6^-$ |
| **Li(Anion-) + 3 EC** | -4738.3 | -4911.1 | -5052.6 |
| **$Li^+$(EC)Anion- + 2 EC** | -4750.1 | -4926.0 | -5062.0 |
| **$Li^+(EC)_2$Anion- + EC** | -4752.2 | -4929.2 | -5067.0 |
| **$Li^+(EC)_3$Anion-** | -4749.1 | -4926.5 | -5064.2 |

Second, the conclusions based on the DFT calculations are in full agreement with the experimental results, where we have observed that each $Li^+$ ion binds to 2.1 ± 0.2 EC molecules, on average. The anion influence is also evident from our experimental data sets. Indeed, the average rotation decay time of the $Li^+$-bound EC complexes slows down with the anion volume sequence, $BF_4^- < ClO_4^- \leq PF_6^-$. The decay time ratio, in contrast, is only slightly larger than two, implying that the effect of the anion on the rotational decay of the complex is only mild, without any evident correlation with the nature of the anion. This behavior is simple to rationalize, by observing that the anion binding energy to the complex is much weaker than that associated to the binding of the cation. The structure of solvated cation/anion weak complexes in Figure 2 lies between those of the contact ion pair and the solvent separated ion pair. The formation of such complexes provides a rational explanation for the ionic conductivity drop of lithium/carbonate electrolyte solutions at high concentrations that is not well correlated with the solution viscosity change[22,23]. Because of the complexation, the charge of solvated $Li^+$ is neutralized by the anion and its mobilty under applied potential can be significantly reduced. At the same time, the weak binding between the solvated cation and anion does not increase the viscosity much, as observed by the rotational dynamics measurments.

Our DFT calculations also show that a bidentate coordination is obtained in the cases of the $Li^+$(EC)(Anion-) and $Li^+(EC)_2$(Anion-) clusters. This is at variance with the case of the $Li^+(EC)_3$(Anion-) complex, where the steric repulsion between the anion and the solvent molecules increases substantially, and a monodentate linkage is therefore observed (see Figures S8-S10 in the SI).

We conclude this Section by describing additional insight based on the DFT vibrational frequency analysis that we have used to extract the vibrational IR spectra. We have considered the clusters indicated in Figure 3, and compared them with the case of the isolated EC molecule. Focusing on the O-C-O stretch band, the experimental absorption frequency peak is located at 1165 $cm^{-1}$ in the case of isolated EC, while the value calculated by DFT is 1074 $cm^{-1}$. In the case of the ($Li^+(EC)_2$Anion-) clusters, in the DFT calculation the O-C-O frequency is blue-shifted in all cases, and the peaks are located around 1150 $cm^{-1}$, with a shift of about 76 $cm^{-1}$. The corresponding blue-shift in the experiments is about two times smaller (around 38 $cm^{-1}$). We note that these discrepancies can be certainly also attributed to the fact that the DFT calculations are performed with (isolated) clusters in the gas phase, while experiments are made in solution (bulk) phases. As a result, the DFT calculations only take into account interactions within the first solvation shell around $Li^+$ ion, disregarding longer-range intermolecular interactions with the condensed bulk phase. These have been demonstrated to be crucial in Ref. 17. Notwithstanding these differences, however, the overall trend in both the experiments and DFT calculations is very similar, further confirming the relevance of the DFT calculations for a full rationalization of the experimental data.

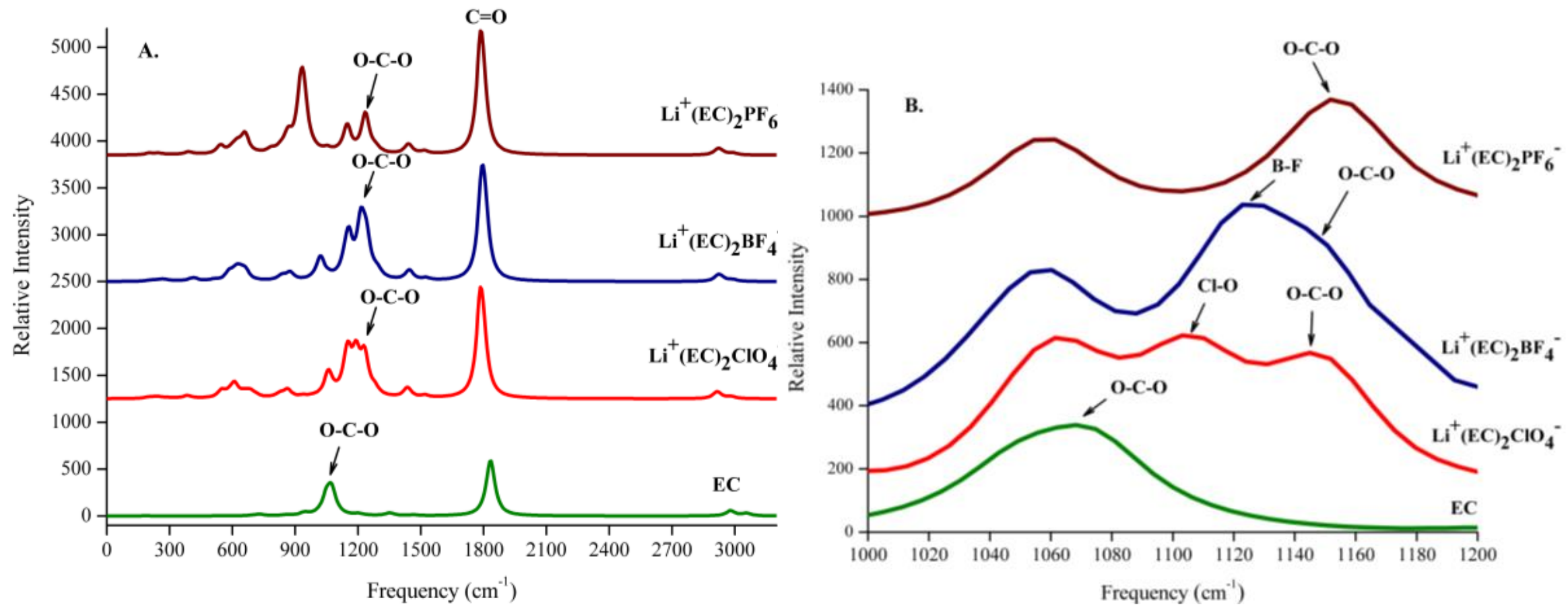

***Figure 3:*** Calculated vibrational IR spectra of isolated EC and of ($Li^+(EC)_2$(Anion-)) optimized clusters (A) and corresponding O-C-O stretching frequency regions (B).

In summary, dynamical spectroscopy methods have been employed to clarify the solvation structure of lithium-salts in EC, a popular Li-ion batteries electrolyte, and provide a *direct* determination of the solvent coordination number around the $Li^+$ ions by measuring the orientational correlation dynamics of the solvent molecules. The presence of the anions, and the salt concentration itself, both have a strong effect on the solvation structure. In particular, we have demonstrated that the local tetrahedral structure around $Li^+$, widely accepted in dilute conditions, is strongly modified in concentrated solutions, leading to a first coordination shell of $Li^+$ comprising two EC molecules only. We have also clarified the elusive role played by the anion, by employing systematic DFT calculations performed for numerous clusters of the type $Li^+(EC)_{1\text{-}3}(Anion^-)$, with $ClO_4^-$, $BF_4^-$ and $PF_6^-$ the considered anions. Based on a detailed analysis of the changes in free energy associated to the transitions between different types of clusters, we have concluded that the formation of complexes of the type $Li^+(EC)_2(Anion^-)$ is the most favorable. These results demonstrate for the first time, to the best of our knowledge, the anion influence on the overall structure of the first solvation shell of the $Li^+$ ion. This information provides deep insight on the solvation structure of non-dilute Lithium salt solutions in organic electrolytes, providing a rational explanation for the ionic conductivity drop of lithium/carbonate electrolyte solutions at high concentrations. It also is, we believe, an important contribution in view of the rational design of electrolytes for battery applications with optimized properties.

## Acknowledgements

This work is supported by the National Natural Science Foundation of China (No. 21373213), the Chinese Academy of Sciences, and the Ministry of Science and Technology. J.R.Z. is supported by NSF (CHE-1503865), the Welch foundation under Award No. C-1752, the Packard Fellowship, the Sloan Fellowship, and PKU. Y. Zhang thanks a Schlumberger future faculty award. The theoretical part of the work (developed by V.P., V.V, S.M and I.S.) was financially supported by the ANR-2011 PRGE002-04 ALIBABA and the DSM-Energie CEA Program.

## Supporting Information

The supporting Information is available free of charge on the ACS Publications website.

Experimental methods, Anisotropy decay of CO stretch in Li salt EC solutions, 2D IR chemical exchange measurements, Theoretical results.

TOC

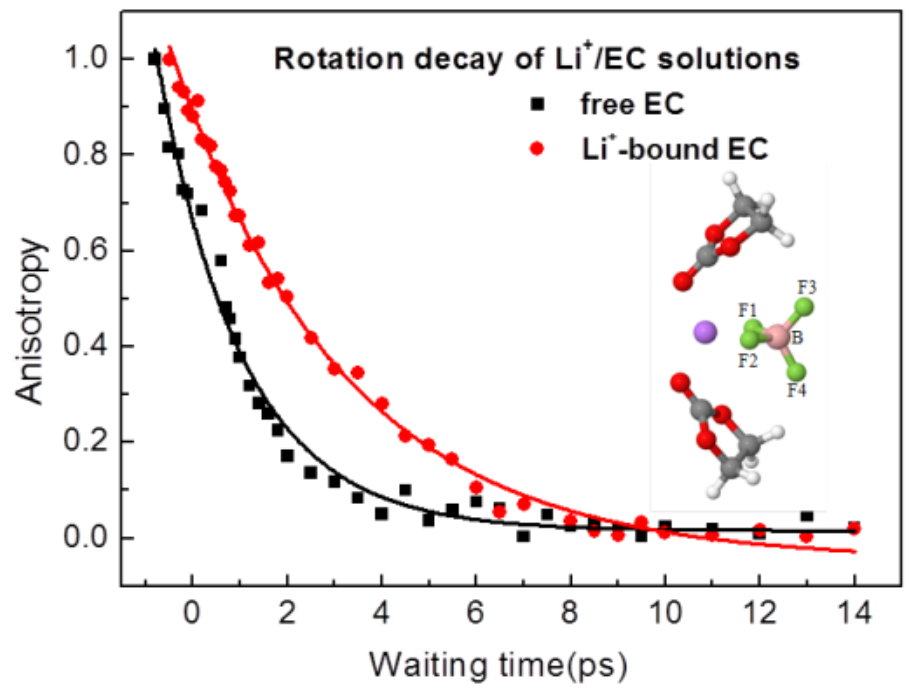

## Supporting information

# The Anion Effect on $Li^+$ Ion Coordination Structure in Ethylene Carbonate solutions.

Bo Jiang[1,3#], Veerapandian Ponnuchamy[4,5#*,], Yuneng Shen[1#], Xueming Yang[1], Kaijun Yuan[1*], Valentina Vetere[5], Stefano Mossa[4], Ioannis Skarmoutsos[4*], Yufan Zhang[3], Junrong Zheng[1,2,3]*,

**Experimental methods**

The laser setup is similar to that described in previous work[1]. A seed light, which generated from an oscillator, is separated to picosecond amplifier and femtosecond amplifier which are synchronized. The picosecond amplifier pumps an OPA to generate ~1.1 ps laser pulse with a bandwidth 10-30 $cm^{-1}$ in a frequency tunable range from 900-4000 $cm^{-1}$ with energy ~30 μJ (5-35 μJ with different frequency) per pulse at the rate of 1 KHz. The femtosecond amplifier pumps an OPA to generate ~60 fs laser pulse with a bandwidth 200 $cm^{-1}$ in a frequency tunable range from 900-4000 $cm^{-1}$ with energy 20 μJ/pulse at the same rate. In nonlinear IR experiments, the picosecond mid-IR pulse is the pump beam (pump power is based on need, the interaction spot varies from 100-500 μm). The fs IR pulse is the probe beam, which is frequency resolved by a spectrograph (resolution is 1-3 $cm^{-1}$ depended on frequency) yielding the probe frequency axis of a multiple-dimensional spectrum. Scanning the pump frequency yields the other frequency axis of the spectrum. Scanning the delay between the pump and the probe provides the time axis. To measure the parallel and perpendicular (relative to pump beam) polarized signal separately, two polarizers are inserted into the probe beam path. The entire setup is controlled by computer. Vibrational lifetimes are obtained from rotation free signal $P_{life}=P_{\parallel}+2\times P_{\perp}$, where $P_{\parallel}$ and $P_{\perp}$ are parallel and perpendicular signal, respectively. Rotational relaxation dynamics are acquired from the time dependent anisotropy $R=(P_{\parallel}-P_{\perp})/(P_{\parallel}+2\times P_{\perp})$. In the analyses of reorientation dynamics, the heat from the CO vibrational relaxation was removed, following the procedure in our previous publication[2]: the heat signal is assumed to grow with time constants slightly slower than the lifetimes of vibrational excitations of which the relaxation generates heat. The maximum amplitude of the heat signal is the transient signal at very long waiting times when most vibrational excitations have relaxed. The time dependent heat signal calculated is then subtracted from the transient signal.

Lithium salts were purchased from Alfa Aesar with >98% purity. Ethylene carbonate was purchased from Acros Organics. The solutions were prepared in glovebox to protect the lithium salt hydrolyzed by water.

The temperature at which measurements were performed is about 40$^{o}$C (313K), since EC is a solid at room temperature. The DFT calculations are performed at 298K, which is a little different to that in experiments. However, the calculations are not very sensitive to temperature. Therefore the temperature difference between theory and experiments is small.

## Anisotropy decay of CO stretch in Li salt EC solutions

### a) Anisotropy decay of CO stretch in pure EC solutions

Figure S2 displays the anisotropy decay data of O-C-O stretch vibrational excitation in the pure EC solution at room temperature. The decay is very fast, with a single exponential decay time constant 1.7±0.1 ps. Generally, there are three possible molecular origins responsible for the signal anisotropy decay.[3,4] One is the molecular rotation. Another one is the nonresonant energy transfer from a donor to a nonresonant energy acceptor and then to another molecule of the same type as original energy donor but with different orientation. For most molecules, this contribution to anisotropy decay is too small to consider. Because it takes a long time (tens or hundreds of picoseconds) for the two-stepped nonresonant energy transfer to occur, which are much slower than a small molecule rotation time (a few picoseconds) in liquids[5]. The third contribution is resonant energy transfer from the energy donor to a randomly oriented acceptor with the same frequency. In our previous work, we measured the resonant energy transfers of CN stretches among different $CH_3CN$ molecules in $CH_3CN$ liquids diluted with $CH_3{}^{13}C^{15}N$. The resonant energy transfer is much slower than 2 ps. Because the transition dipole moment of the O-C-O stretch of EC is much smaller than that of CN stretch and the average molecular distances in EC and $CH_3CN$ are similar. According to the dipole/dipole interaction, the resonant energy transfer of O-C-O among different EC molecules must be slower than that of CN stretch. In other words, it must be much slower than 2 ps. Therefore, the experimentally observed anisotropy decay 1.7 ps is mainly contributed by the molecular rotation.

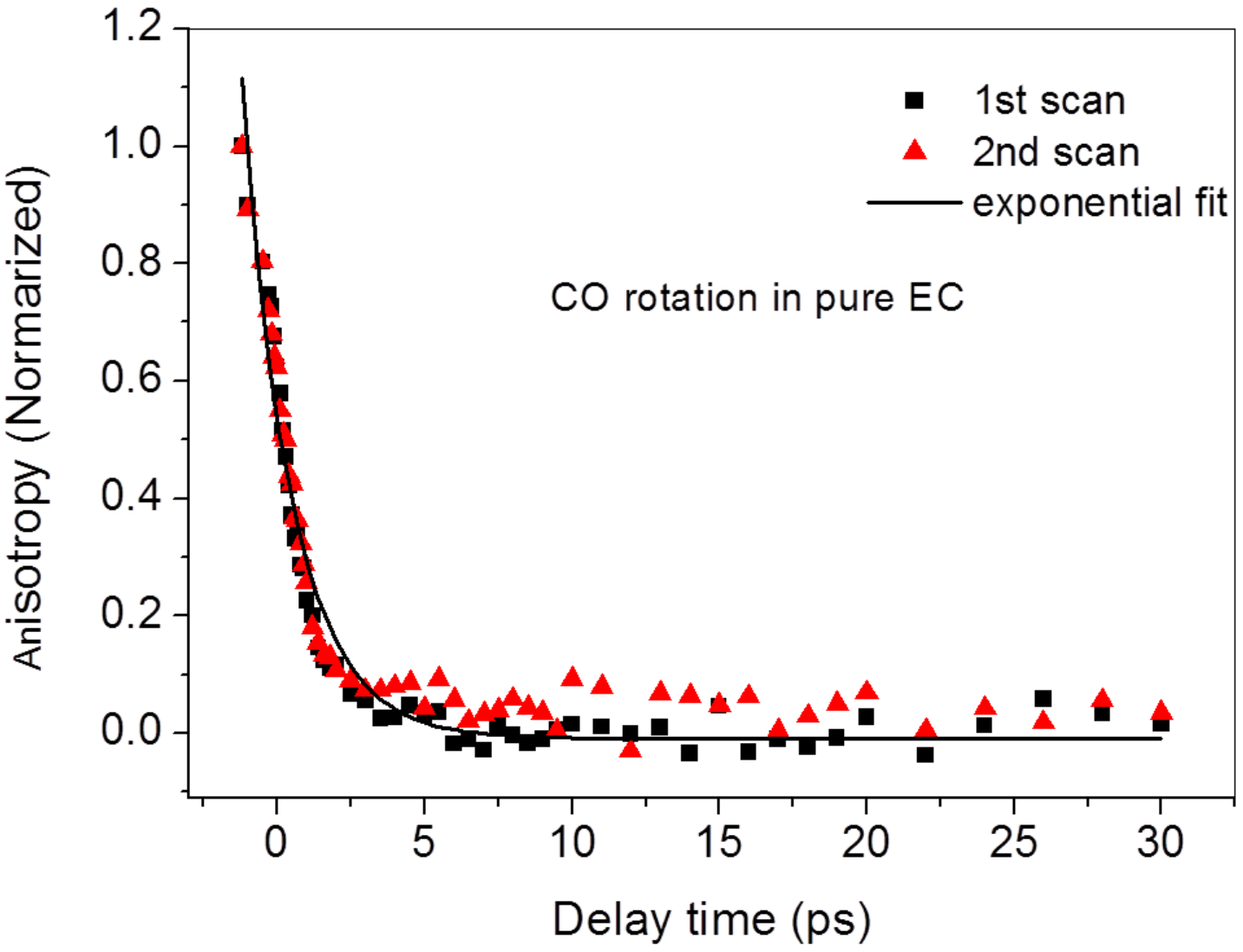

***Figure S1.*** *Anisotropy decay curves of the CO stretch vibrational excitation signal of EC in pure EC. Dots are data, and the curves are fits of a single exponential with a time of constant ~1.7 ps.*

b) **Anisotropy decay of CO stretch in $LiClO_4$ / EC solutions**

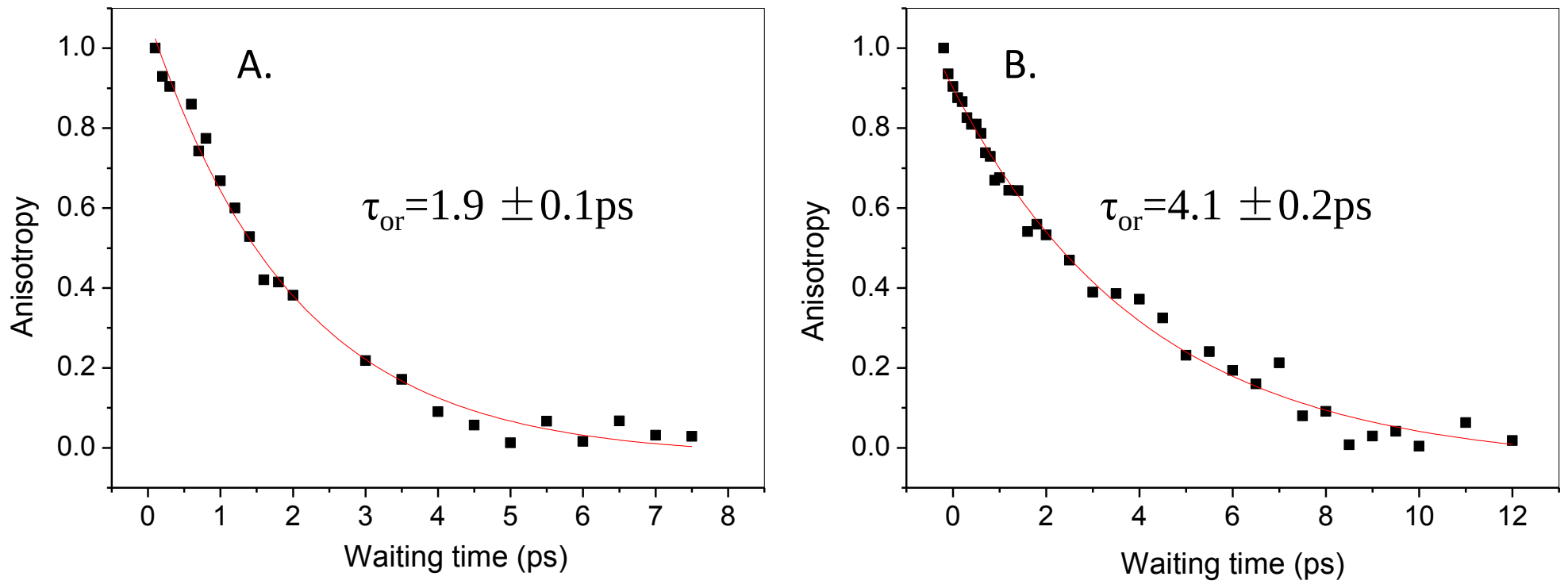

***Figure S2.*** *Time dependent anisotropies of CO stretch of free EC (1165 cm$^{-1}$) (A) and Li$^{+}$-bound EC (1203 cm$^{-1}$)(B) in a 1/5 mol% $LiClO_4$/EC solution.*

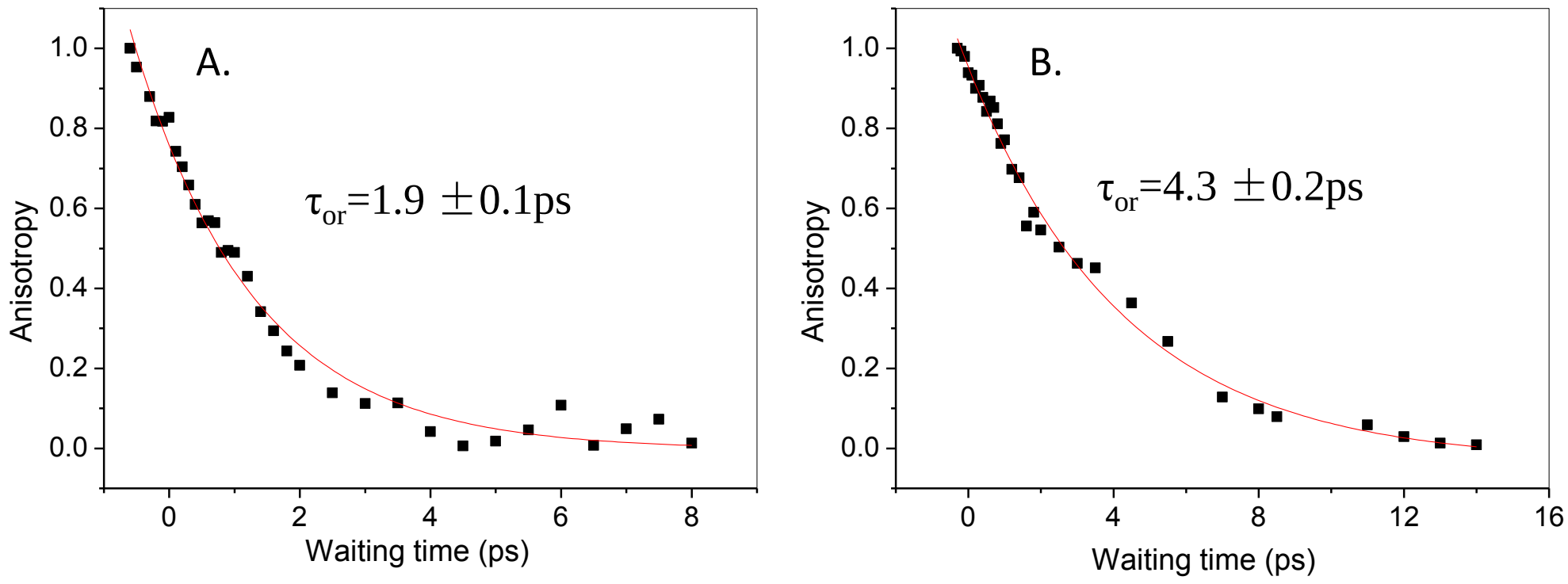

***Figure S3.*** *Time dependent anisotropies of CO stretch of free EC (1165 $cm^{-1}$) (A) and $Li^{+}$-bound EC (1203 $cm^{-1}$)(B) in a 1/10 mol% $LiClO_4$/EC solution.*

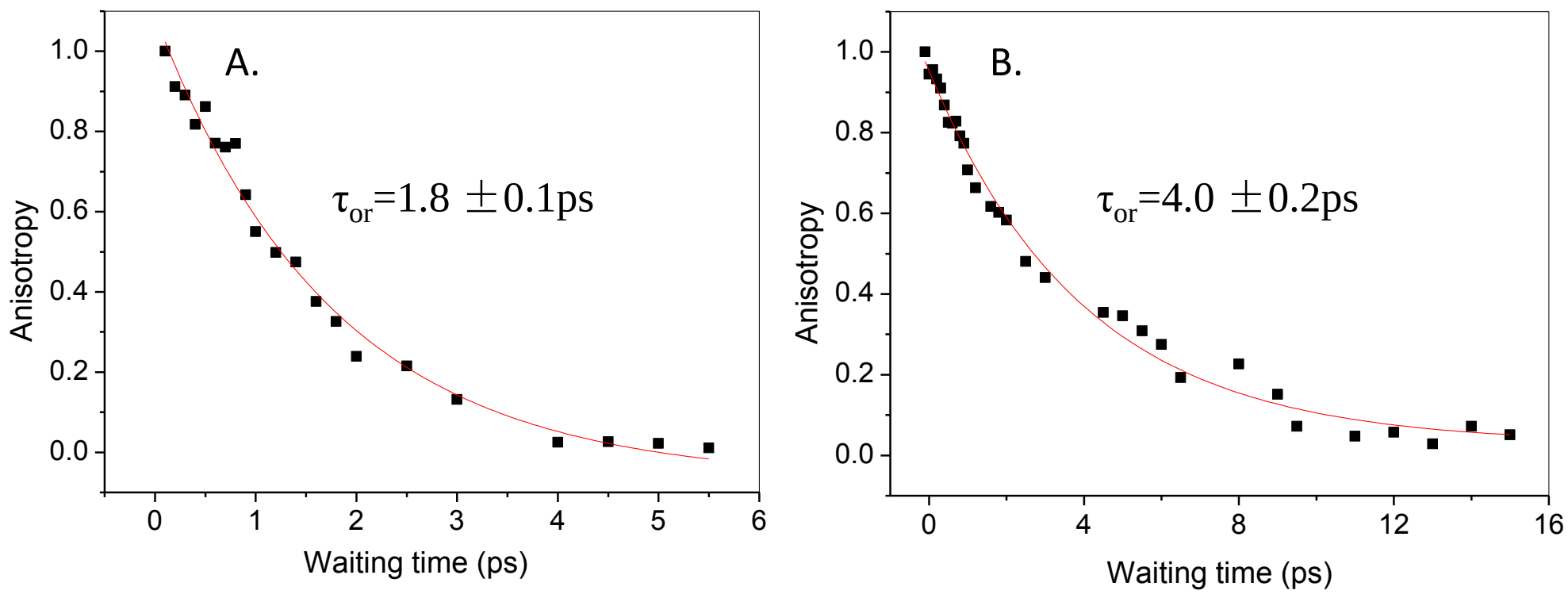

***Figure S4.*** *Time dependent anisotropies of CO stretch of free EC (1165 $cm^{-1}$) (A) and $Li^{+}$-bound EC (1203 $cm^{-1}$)(B) in a 1/20 mol% $LiClO_4$/EC solution.*

**c) Anisotropy decay of CO stretch in $LiPF_6$ / EC solutions**

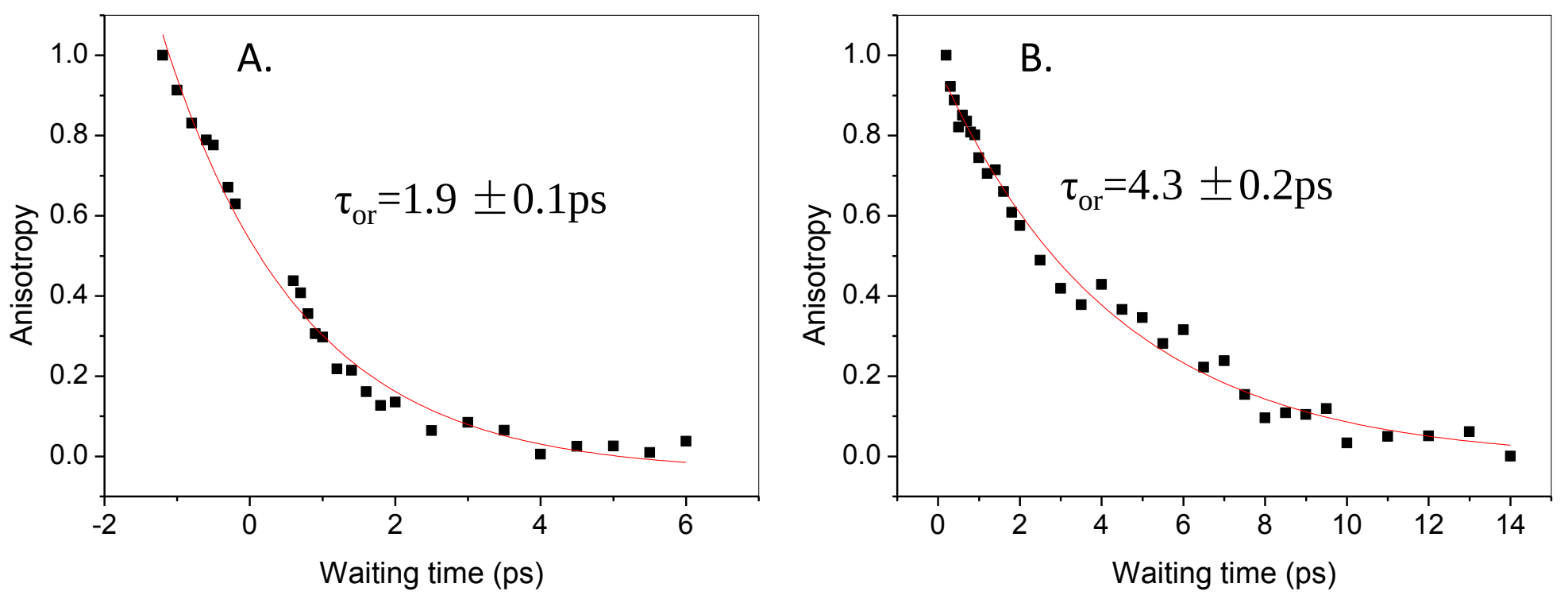

***Figure S5.*** *Time dependent anisotropies of CO stretch of free EC (1165 $cm^{-1}$) (A) and*

*Li⁺-bound EC (1203 cm$^{-1}$)(B) in a 1/10 mol% $LiPF_6$/EC solution.*

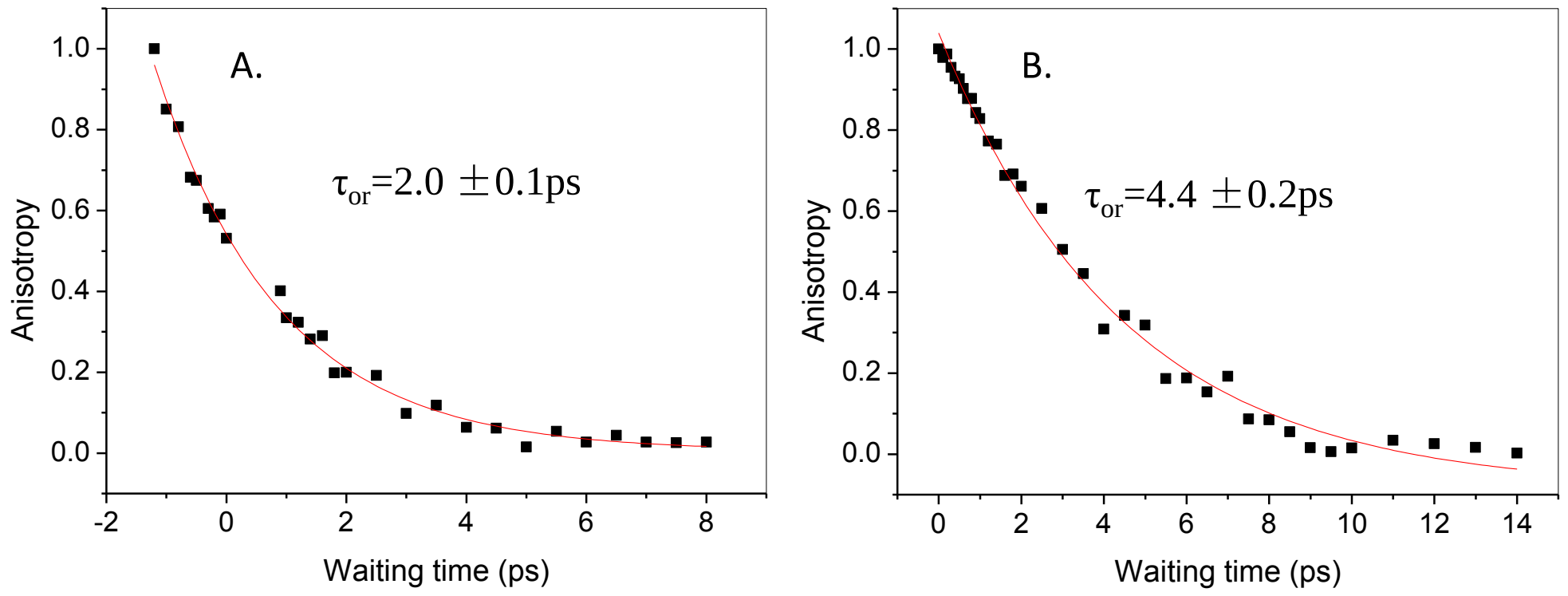

***Figure S6.*** *Time dependent anisotropies of CO stretch of free EC (1165 cm$^{-1}$) (A) and $Li^+$-bound EC (1203 cm$^{-1}$)(B) in a 1/20 mol% $LiPF_6$/EC solution.*

**2D IR chemical exchange measurements**

The 2D IR chemical exchange measurements are shown in Figure S7. The red peak (1165 cm$^{-1}$, 1165 cm$^{-1}$) and the blue peak (1165 cm$^{-1}$, 1152 cm$^{-1}$) belong to the 0-1 and 1-2 transition peaks of CO stretch of free EC molecules. While the red peak (1203 cm$^{-1}$, 1203 cm$^{-1}$) and the blue peak (1203 cm$^{-1}$, 1185 cm$^{-1}$) belong to the 0-1 and 1-2 transition peaks of CO stretch of $Li^+$-bound EC. If the $Li^+$-bound EC complex dissociate to $Li^+$ and free EC molecules, there will be a blue peak grows up at the position (1203, 1165 cm$^{-1}$) in 2D IR spectra in a few picoseconds after the excitation of CO stretch. In the same way, a new red cross peak will grows up at position (1165, 1203 cm$^{-1}$) if free EC molecules binding with $Li^+$ after a long waiting time. It seems that the blue peak extends up to position (1203, 1165 cm$^{-1}$). Due to the very short vibrational lifetime of CO stretch (~2 ps), the energy releases to environment quickly and can cause a strong heat effect in position (1203, 1165 cm$^{-1}$). Thus it is not clear the source of the signal at position (1203, 1165 cm$^{-1}$) immediately. In our previous work, we have estimated the lifetime of $Li^+$-bound $CH_3CN$ complex by calculating the formation enthalpy. The formation enthalpy of Li-N in $Li^+$-$CH_3CN$ complex is about 47 kcal/mol and the estimated lifetime is longer than 19 ps**Erreur ! Signet non défini.**, while that of Li-O in $Li^+$-EC is about 59 kcal/mol, which suggests that the lifetime of complex $Li^+$-bound EC should be longer than $Li^+$-$CH_3CN$.

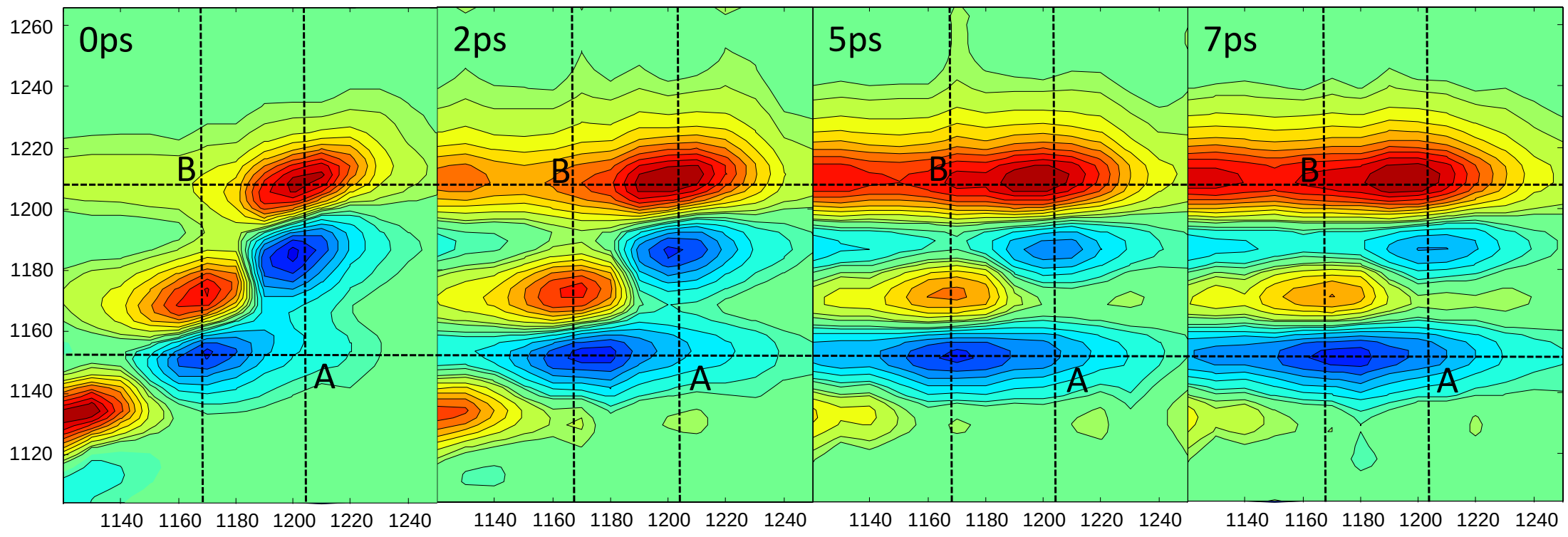

***Figure S7.*** *Waiting time dependent 2D IR spectra of $LiClO_4$/EC 1/10 mol% solution.*

**Relative conductivity**

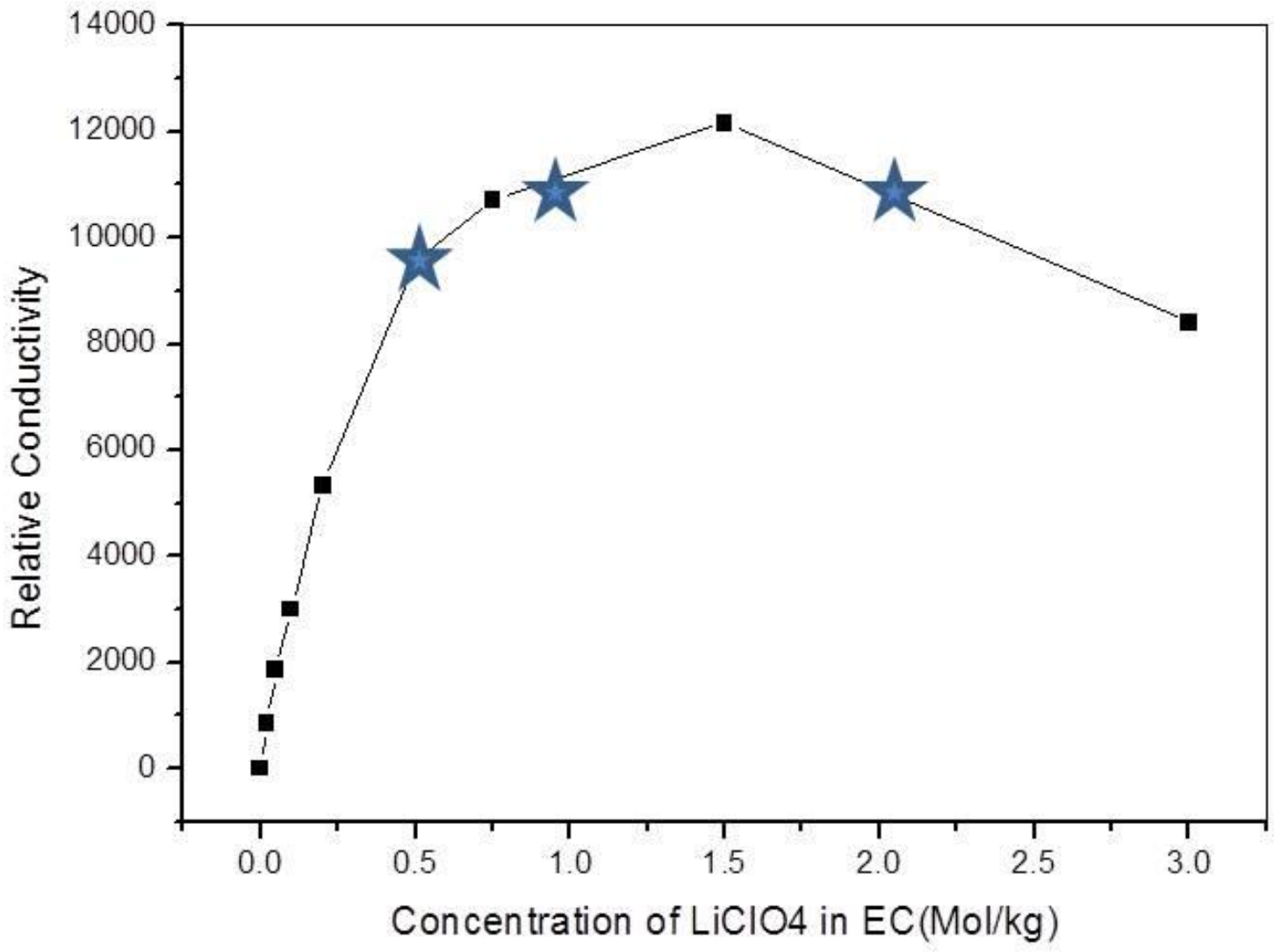

***Figure S8.*** *Relative conductivity of $LiClO_4$/EC solutions.*

## Theoretical Results

### a) Structural parameters of optimized $Li^+(EC)_{1\text{-}3}BF_4^-$ clusters

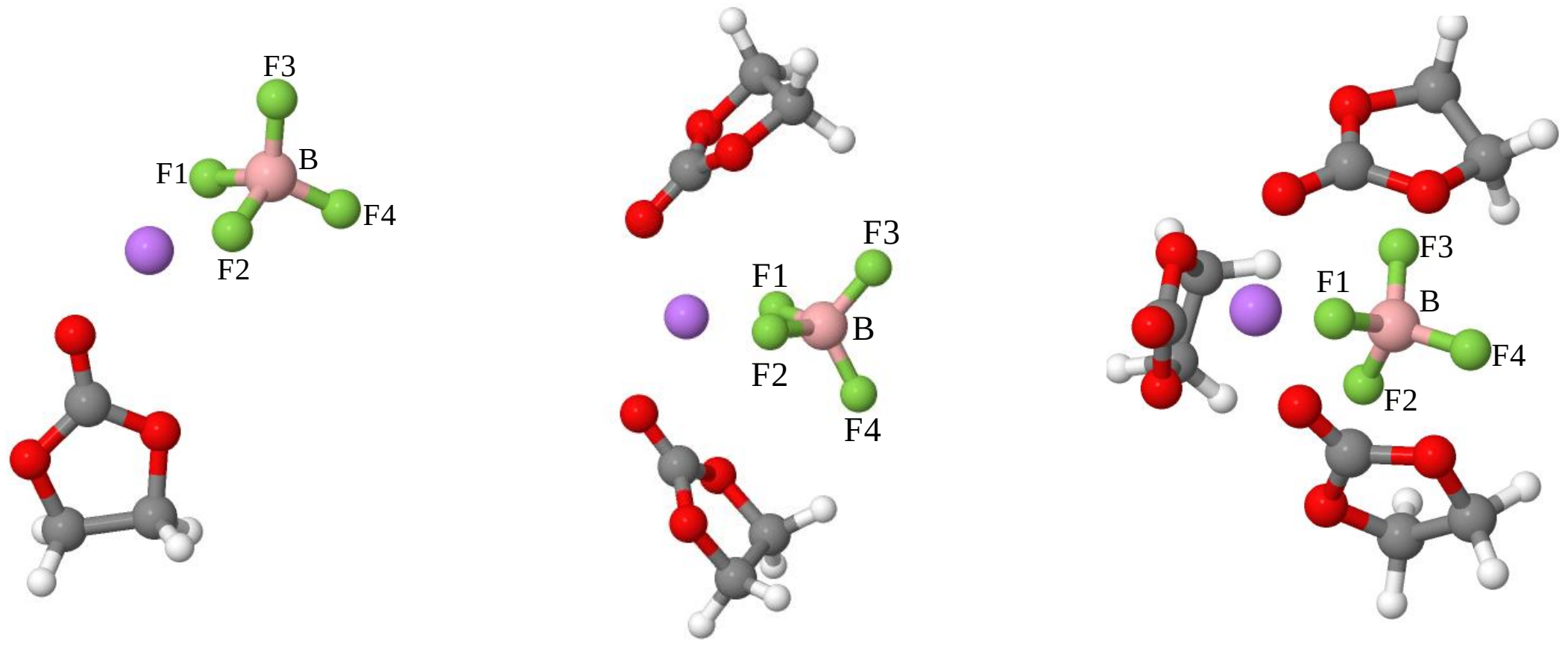

***Figure S9.*** *The optimized geometries of (a) $Li^+(EC)BF_4^-$, (b) $Li^+(EC)_2BF_4^-$ and (c) $Li^+(EC)_3BF_4^-$*

### b) The optimized geometries of $Li^+(EC)_{1\text{-}3}ClO_4^-$ clusters

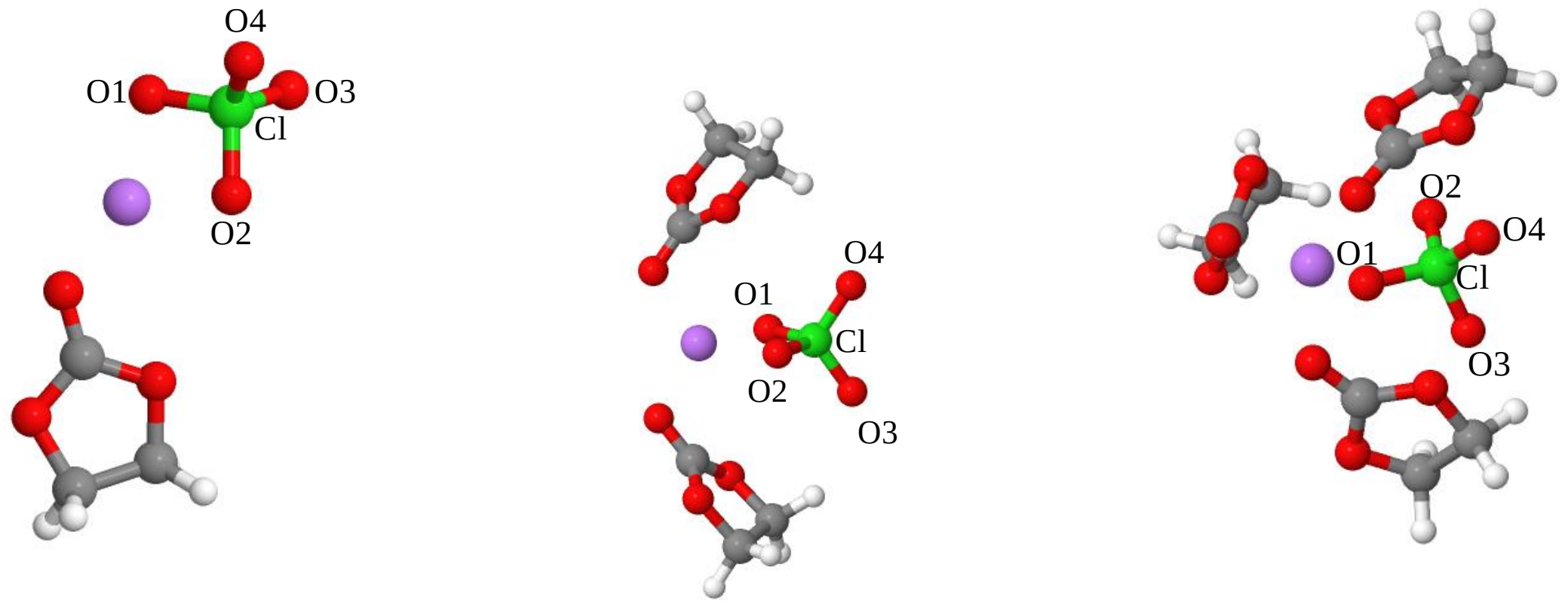

(a) (b) (c)

***Figure S10.*** *The optimized geometries of (a) $Li^+(EC)ClO_4^-$, (b) $Li^+(EC)_2ClO_4^-$ and (c) $Li^+(EC)_3ClO_4^-$*

**c) The optimized geometries of $Li^+(EC)_{1-3}\ PF_6^-$ clusters**

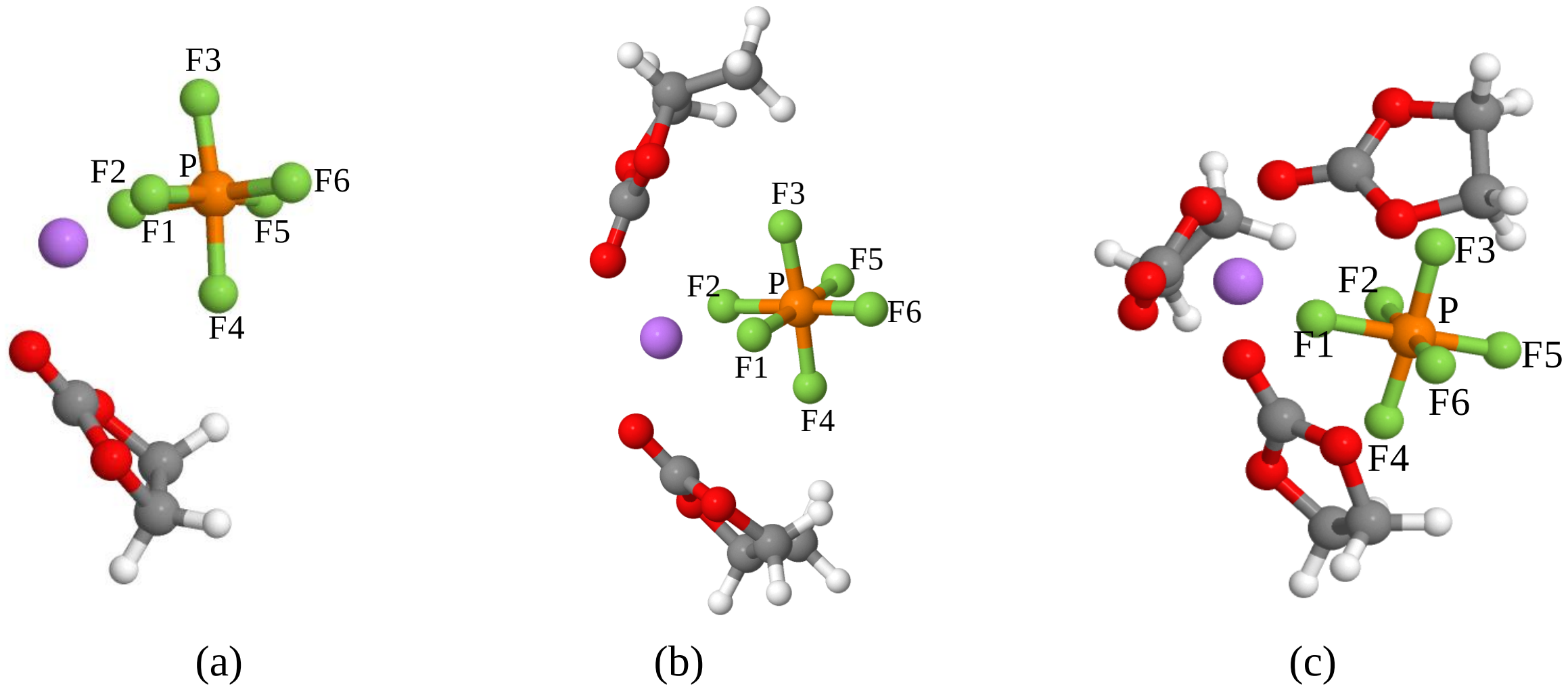

***Figure S11.*** *The optimized geometries of (a) $Li^+(EC)PF_6^-$, (b) $Li^+(EC)_2PF_6^-$ and (c) $Li^+(EC)_3PF_6^-$*

**<u>Optimized geometry Coordinates</u>**

**$Li^+(EC)(BF_4^-)$**

| | | | |
|---|---|---|---|
| H | -0.646754 | -1.551913 | 1.544761 |
| H | -0.454901 | -1.972528 | -0.194034 |
| C | -1.893754 | 0.491751 | -0.438340 |
| O | -0.621273 | 0.928131 | -0.525700 |
| C | 0.211535 | 0.072884 | 0.308637 |
| C | -0.663457 | -1.164952 | 0.520573 |
| O | -2.001008 | -0.663371 | 0.248153 |
| O | -2.845242 | 1.080072 | -0.927592 |
| H | 0.424049 | 0.615624 | 1.239710 |
| H | 1.141689 | -0.129189 | -0.232738 |
| Li | -4.600967 | 0.351045 | -0.921490 |
| B | -6.022044 | -1.461784 | -0.383242 |
| F | -5.758141 | -0.209137 | 0.369588 |
| F | -5.479318 | -2.532172 | 0.293186 |
| F | -7.357625 | -1.565678 | -0.678681 |
| F | -5.240483 | -1.205704 | -1.619935 |

**$Li^+(EC)_2(BF_4^-)$**

| | | | |
|---|---|---|---|
| H | 0.562997 | -0.852648 | 1.062320 |
| H | 0.659864 | 0.945446 | 1.014018 |
| C | -1.774548 | 0.494281 | -0.462982 |
| O | -2.226632 | 0.740563 | 0.783900 |
| C | -1.280228 | 0.166069 | 1.727950 |
| C | 0.001086 | 0.073111 | 0.898386 |
| O | -0.493561 | 0.063153 | -0.469246 |
| O | -2.441187 | 0.647915 | -1.472750 |
| H | -1.674322 | -0.814764 | 2.024269 |
| H | -1.204045 | 0.839893 | 2.587822 |
| Li | -4.012833 | -0.487095 | -1.669674 |
| B | -4.082016 | -2.333150 | -0.001602 |
| F | -3.161984 | -2.370952 | 1.045605 |
| F | -4.972977 | -3.402160 | 0.020849 |
| F | -4.814502 | -1.073594 | 0.018928 |
| F | -3.378374 | -2.272287 | -1.273301 |
| H | -7.669664 | -3.916429 | -2.838059 |
| H | -7.012326 | -3.466572 | -1.215817 |
| C | -6.464544 | -1.049843 | -2.558063 |
| O | -7.694853 | -0.776038 | -2.071036 |
| C | -8.460778 | -2.012188 | -2.037485 |
| C | -7.385094 | -3.090480 | -2.177654 |
| O | -6.291668 | -2.364527 | -2.806019 |
| O | -5.607652 | -0.203269 | -2.752852 |
| H | -9.170665 | -1.993450 | -2.876439 |
| H | -8.999855 | -2.051196 | -1.084802 |

**$Li^+(EC)_3(BF_4^-)$**

| | | | |
|---|---|---|---|
| H | -0.456296 | -1.356016 | 1.130830 |
| H | 0.116250 | 0.325309 | 1.457302 |
| C | -2.077589 | 0.685107 | -0.470686 |
| O | -2.725775 | 0.841410 | 0.702939 |
| C | -2.064608 | 0.030804 | 1.712525 |
| C | -0.707531 | -0.291644 | 1.073188 |
| O | -0.899526 | 0.043637 | -0.326638 |
| O | -2.502042 | 1.087757 | -1.541003 |
| H | -2.680735 | -0.862002 | 1.876362 |
| H | -1.988126 | 0.627817 | 2.628325 |
| Li | -4.028105 | 0.096541 | -2.259646 |
| B | -3.844853 | -2.614292 | -0.147933 |
| F | -2.486605 | -2.884521 | 0.111112 |
| F | -4.468365 | -3.708640 | -0.779673 |
| F | -4.525162 | -2.255814 | 1.033104 |
| F | -3.906231 | -1.473253 | -1.064461 |
| H | -8.453105 | -2.502311 | -0.788429 |
| H | -6.748680 | -2.882251 | -0.324710 |
| C | -6.487805 | -0.123275 | -1.093005 |
| O | -6.787567 | 0.043575 | 0.212363 |
| C | -7.566877 | -1.092314 | 0.671076 |
| C | -7.482809 | -2.085206 | -0.495909 |
| O | -6.996170 | -1.268456 | -1.595370 |
| O | -5.840143 | 0.677575 | -1.745736 |
| H | -8.587412 | -0.740605 | 0.875164 |
| H | -7.097637 | -1.473609 | 1.584221 |
| H | -1.245687 | -4.231240 | -3.657702 |
| H | -2.134913 | -3.921353 | -2.115716 |
| C | -3.423760 | -2.027979 | -3.844399 |
| O | -4.216663 | -3.097664 | -4.063564 |
| C | -3.450877 | -4.311118 | -3.841188 |
| C | -2.149106 | -3.805464 | -3.206693 |
| O | -2.167519 | -2.381198 | -3.497754 |
| O | -3.798965 | -0.873160 | -3.958482 |
| H | -3.303639 | -4.799713 | -4.814134 |
| H | -4.027940 | -4.952454 | -3.166676 |

**$Li^+(EC)(ClO_4^-)$**

| | | | |
|---|---|---|---|
| H | 1.771517 | 3.102887 | -0.492925 |
| H | 2.343348 | 2.153753 | 0.924770 |
| C | 1.000246 | 0.094265 | -0.560526 |
| O | 2.265278 | -0.044707 | -1.006315 |
| C | 2.891935 | 1.268725 | -1.022563 |
| C | 1.983928 | 2.098460 | -0.111754 |
| O | 0.742090 | 1.341245 | -0.113868 |
| O | 0.184565 | -0.812311 | -0.560826 |
| H | 2.901212 | 1.624240 | -2.061853 |
| H | 3.916071 | 1.162661 | -0.649625 |
| Li | -1.661430 | -0.767448 | -0.139679 |
| Cl | -3.967341 | -0.182436 | 0.401657 |
| O | -4.386121 | 0.326513 | 1.714146 |
| O | -3.261955 | -1.537032 | 0.558259 |
| O | -5.087793 | -0.298413 | -0.539898 |
| O | -2.868432 | 0.713447 | -0.187293 |

**$Li^+(EC)_2(ClO_4^-)$**

| | | | |
|---|---|---|---|
| H | -0.139262 | 3.330136 | -0.869528 |
| H | -0.590564 | 2.342890 | 0.571267 |
| C | 0.943994 | 0.403740 | -0.795045 |
| O | 1.904251 | 0.969614 | -0.031109 |
| C | 1.580296 | 2.376541 | 0.145003 |
| C | 0.109066 | 2.449120 | -0.268352 |
| O | -0.041203 | 1.271483 | -1.107585 |
| O | 0.966660 | -0.758280 | -1.165055 |
| H | 2.244456 | 2.961126 | -0.507039 |
| H | 1.754912 | 2.635990 | 1.194729 |
| Li | -0.620355 | -1.760328 | -0.645578 |
| Cl | -2.409125 | -0.625638 | 0.863631 |
| O | -2.661148 | 0.828987 | 0.930507 |
| O | -0.959978 | -0.918603 | 1.218907 |
| O | -3.322353 | -1.366814 | 1.759443 |
| O | -2.562396 | -1.129513 | -0.563289 |
| H | -4.702245 | -5.250924 | -0.581010 |
| H | -3.452999 | -6.502155 | -0.241580 |
| C | -1.794348 | -4.176128 | -0.608771 |
| O | -1.943176 | -4.222310 | 0.731964 |
| C | -3.285353 | -4.699497 | 1.023175 |
| C | -3.666834 | -5.424375 | -0.268763 |
| O | -2.790189 | -4.818347 | -1.258624 |
| O | -0.864745 | -3.620315 | -1.168977 |
| H | -3.905981 | -3.818512 | 1.234339 |
| H | -3.231990 | -5.354234 | 1.899396 |

**$Li^+(EC)_3(ClO_4^-)$**

| | | | |
|---|---|---|---|
| H | 0.913186 | 3.624296 | -0.172762 |
| H | -0.670202 | 2.751696 | -0.182074 |
| C | 1.136620 | 0.549947 | -0.532601 |
| O | 1.079491 | 0.582413 | 0.815445 |
| C | 0.775634 | 1.932270 | 1.251801 |
| C | 0.414881 | 2.655814 | -0.051870 |
| O | 0.908430 | 1.762007 | -1.086064 |
| O | 1.379246 | -0.456338 | -1.175584 |
| H | 1.670112 | 2.336423 | 1.745244 |
| H | -0.066057 | 1.877101 | 1.950739 |
| Li | -0.038722 | -1.828591 | -1.103393 |
| Cl | -2.463356 | -0.305000 | 0.561381 |
| O | -2.481358 | 1.164248 | 0.781467 |
| O | -1.891979 | -1.002258 | 1.748584 |
| O | -3.835071 | -0.810750 | 0.275587 |
| O | -1.585636 | -0.588757 | -0.650296 |
| H | -4.019664 | 0.124461 | -3.844070 |
| H | -3.932602 | -1.224021 | -5.034393 |
| C | -1.890094 | -2.011363 | -3.137647 |
| O | -2.971054 | -2.705018 | -2.714984 |
| C | -4.149897 | -1.882712 | -2.936084 |
| C | -3.680332 | -0.903459 | -4.013320 |
| O | -2.235700 | -0.931029 | -3.877660 |
| O | -0.738842 | -2.327445 | -2.900305 |
| H | -4.390390 | -1.391400 | -1.983532 |
| H | -4.966822 | -2.536645 | -3.260591 |
| H | -1.702870 | -5.483845 | 2.935976 |
| H | -2.045763 | -3.885672 | 3.691533 |
| C | -0.705061 | -3.619963 | 0.931386 |
| O | -1.954955 | -3.872678 | 0.484798 |
| C | -2.817152 | -4.086649 | 1.632818 |
| C | -1.822678 | -4.406507 | 2.753869 |
| O | -0.562541 | -3.896508 | 2.245552 |
| O | 0.202227 | -3.218257 | 0.225271 |
| H | -3.495957 | -4.914093 | 1.398113 |
| H | -3.374574 | -3.157493 | 1.805087 |

**$Li^{+}(EC)(PF_6^{-})$**

| | | | |
|---|---|---|---|
| F | -7.659580 | 2.825268 | 3.847032 |
| P | -8.695486 | 1.725277 | 4.528450 |
| Li | -6.674005 | 0.296703 | 3.501835 |
| F | -8.432386 | 0.754984 | 3.127419 |
| F | -7.291655 | 0.924189 | 5.135448 |
| F | -8.787273 | 2.600282 | 5.874211 |
| F | -9.596246 | 0.536651 | 5.142658 |
| F | -9.955599 | 2.427200 | 3.824612 |
| O | -4.918062 | 3.136480 | 1.916995 |
| O | -5.147375 | 1.018633 | 2.682771 |
| H | -3.811967 | 4.567726 | 4.594397 |
| H | -5.605691 | 4.462244 | 4.411432 |
| H | -3.478977 | 4.610024 | 2.174725 |
| H | -5.186203 | 5.182977 | 2.126144 |
| C | -4.899895 | 2.201222 | 2.883006 |
| O | -4.568872 | 2.701555 | 4.087157 |
| C | -4.634736 | 4.153892 | 4.002440 |
| C | -4.508865 | 4.407566 | 2.498990 |

**$Li^+(EC)_2(PF_6^-)$**

| | | | |
|---|---|---|---|
| F | -8.625894 | 1.399906 | 3.082227 |
| F | -9.341046 | 0.730064 | 5.201466 |
| Li | -7.141918 | 0.147470 | 2.907606 |
| F | -9.671476 | 2.936804 | 4.522940 |
| F | -8.075043 | 2.427008 | 6.180228 |
| F | -7.067551 | 0.901340 | 4.701838 |
| P | -8.393255 | 1.964118 | 4.667877 |
| F | -7.393108 | 3.122120 | 4.061541 |
| O | -5.347022 | 3.122863 | 1.444933 |
| O | -5.758307 | 0.929601 | 1.830787 |
| H | -3.225552 | 3.701380 | 3.804380 |
| H | -4.969519 | 3.887325 | 4.237420 |
| H | -3.648036 | 4.312321 | 1.483753 |
| H | -5.149713 | 5.069217 | 2.129086 |
| C | -5.237108 | 1.983080 | 2.158258 |
| O | -4.471551 | 2.135131 | 3.258528 |
| C | -4.258315 | 3.557112 | 3.469643 |
| C | -4.550668 | 4.153124 | 2.089947 |
| O | -10.077253 | -1.505917 | 3.040183 |
| O | -7.813231 | -1.647278 | 3.021725 |
| H | -10.573434 | -2.425539 | 6.080196 |
| H | -10.708658 | -3.729459 | 4.844962 |
| H | -11.292398 | -0.720714 | 4.517662 |
| H | -11.992953 | -2.094219 | 3.576346 |
| C | -8.886486 | -1.874035 | 3.555995 |
| O | -9.013167 | -2.541347 | 4.721431 |
| C | -10.426732 | -2.682863 | 5.025719 |
| C | -11.093956 | -1.698401 | 4.060727 |

**$Li^+(EC)_3(PF_6^-)$**

| | | | |
|---|---|---|---|
| H | -1.050973 | 0.010980 | 5.806580 |
| H | -0.765596 | 0.310430 | 4.048251 |
| C | -2.976374 | 2.056332 | 4.470854 |
| O | -1.960073 | 2.907688 | 4.737727 |
| C | -0.843740 | 2.147761 | 5.271293 |
| C | -1.223012 | 0.694688 | 4.968025 |
| O | -2.656413 | 0.774374 | 4.740001 |
| O | -4.063244 | 2.416592 | 4.054407 |
| H | -0.769406 | 2.364792 | 6.346199 |
| H | 0.065842 | 2.472787 | 4.754303 |
| Li | -4.863299 | 1.682843 | 2.430840 |
| H | -5.124038 | -3.482698 | 0.727056 |
| H | -6.900997 | -3.217727 | 0.597325 |
| C | -6.036643 | -0.683247 | 1.687586 |
| O | -6.077631 | -0.428990 | 0.362051 |
| C | -5.680008 | -1.635474 | -0.346182 |
| C | -5.920187 | -2.730975 | 0.694877 |
| O | -5.903884 | -2.003240 | 1.952429 |
| O | -6.118908 | 0.172936 | 2.550358 |
| H | -4.622470 | -1.525305 | -0.617198 |
| H | -6.308246 | -1.728719 | -1.238608 |
| H | -2.778239 | 4.370026 | -1.950209 |
| H | -3.604106 | 5.820431 | -1.273578 |
| C | -4.421238 | 3.695795 | 0.559853 |
| O | -3.350622 | 4.215295 | 1.196808 |
| C | -2.405367 | 4.687901 | 0.202095 |
| C | -3.262337 | 4.797036 | -1.064996 |
| O | -4.428884 | 3.989581 | -0.758601 |
| O | -5.296699 | 3.052646 | 1.111679 |
| H | -1.610644 | 3.936905 | 0.118415 |
| H | -2.003985 | 5.648061 | 0.544786 |
| F | -1.418695 | -0.583233 | 1.950293 |
| F | -2.525214 | -0.714870 | -0.088696 |
| F | -2.450943 | 1.602587 | -0.275415 |
| P | -1.898172 | 0.499682 | 0.818640 |
| F | -0.472010 | 0.405296 | 0.052187 |
| F | -1.346925 | 1.733774 | 1.770583 |
| F | -3.396420 | 0.617230 | 1.621757 |

***Table S1**:* Enthalpy, entropy and Gibbs free energy of Clusters $Li^+(EC)_{1-3}$Anion- (where Anion- = $ClO_4^-$, $BF_4^-$, $PF_6^-$) (kcal/mol), obtained when using the COSMO method.

| | H | S | G |
|---|---|---|---|
| $Li^+(EC)ClO_4^-$ + 2 EC | -4712.26 | 257.525 | -4789.04 |
| $Li^+(EC)_2ClO_4^-$ + 1 EC | -4720.94 | 230.689 | -4789.72 |
| $Li^+(EC)_3ClO_4^-$ | -4724.884 | 182.462 | -4779.29 |

| | H | S | G |
|---|---|---|---|
| $Li^+(EC)BF_4^-$ + 2 EC | -4890.23 | 247.27 | -4963.96 |
| $Li^+(EC)_2BF_4^-$ + 1 EC | -4898.3 | 220.465 | -4964.03 |
| $Li^+(EC)_3BF_4^-$ | -4900.55 | 188.587 | -4956.77 |

| | H | S | G |
|---|---|---|---|
| $Li^+(EC)PF_6^-$ + 2 EC | -5024.04 | 267.365 | -5103.75 |
| $Li^+(EC)_2PF_6^-$ + 1 EC | -5034.344 | 238.185 | -5105.36 |
| $Li^+(EC)_3PF_6^-$ | -5037.222 | 197.201 | -5096.02 |